\documentclass[useAMS,usenatbib,longnamesfirst,usedcolumn]{mn2e} % Normal MNRAS format
\pdfoutput=1	% Output PDF only
\usepackage{epsfig}	% Allows included graphics
\usepackage{times}	% Smaller font size
\usepackage{subfigure}	% Allows side-by-side figures
\usepackage[a4paper=true,pagebackref,
  pdftitle={X-ray modelling of galaxy cluster gas and mass profiles},
  pdfauthor={Alastair J.~R. Sanderson \& Trevor J. Ponman},
  pdfsubject={MNRAS},
  pdfkeywords={Chandra, cool-core, cluster, model},
  pdfproducer={LaTeX with hyperref},
  pdfcreator={LaTeX},
  pdfdisplaydoctitle=true,
  colorlinks=true,
  citecolor=blue,
  linkcolor=blue,
  urlcolor=blue]{hyperref}

\title[Modelling cluster gas and mass profiles]{X-ray modelling 
of galaxy cluster gas and mass profiles}

\author[A.~J.~R. Sanderson and T.~J. Ponman]
       {Alastair J.~R. Sanderson\thanks{E-mail: ajrs@star.sr.bham.ac.uk}
        and Trevor J. Ponman\\
 School of Physics and Astronomy, University of
        Birmingham, Edgbaston, Birmingham B15 2TT, UK \\
       \\}
 \date{Accepted 2009 October 15.
      Received 2009 September 15 ($svn$ $Revision: 60 $)}

\pagerange{\pageref{firstpage}--\pageref{lastpage}}
\pubyear{2009}

\shortcites{boehringer07,clarke09,croston08,fabian06a,finoguenov06,
 ikebe02,rasia06,richard09,vikhlinin07,vikhlinin06}

\defcitealias{ascasibar08}{AD08}
\defcitealias{san09}{SOP09}
\newcommand{\rmsub}[2]{\ensuremath{#1_{\mathrm{#2}}}} % correct unit labels
\newcommand{\srel}[2]{\mbox{\ensuremath{#1 - #2}}} % scaling relation def.
\newcommand{\Chandra}{\emph{Chandra}}
\newcommand{\chisq}{\ensuremath{\chi^2}}

\newcommand{\km}{\ensuremath{\mbox{~km}}}
\newcommand{\kmpspMpc}{\ensuremath{\km \ps \pMpc\,}}
\newcommand{\Mpc}{\ensuremath{\mbox{~Mpc}}}
\newcommand{\MT}{\srel{M}{\TX}}

\newcommand{\pMpc}{\ensuremath{\Mpc^{-1}}}
\newcommand{\ps}{\ensuremath{\s^{-1}}}
\newcommand{\rhogas}{\rmsub{\rho}{gas}}
\newcommand{\Rproject}{\textsc{r}}
\newcommand{\rfiveh}{\rmsub{r}{500}}
\newcommand{\s}{\ensuremath{\mbox{~s}}}
\newcommand{\TX}{\rmsub{T}{X}}
\newcommand{\XMM}{\emph{XMM-Newton}}
\begin{document}

\maketitle

\label{firstpage}

\begin{abstract}
  
 \noindent We present a parametric analysis of the intracluster medium and
 gravitating mass distribution of a statistical sample of 20 galaxy
 clusters using the phenomenological cluster model of
 \citeauthor{ascasibar08}. We describe an effective scheme for the
 estimation of errors on model parameters and derived quantities using
 bootstrap resampling. We find that the model provides a good description
 of the data in all cases and we quantify the mean fractional intrinsic
 scatter about the best-fit density and temperature profiles, finding this
 to have median values across the sample of 2 and 5 per cent,
 respectively. In addition, we demonstrate good agreement between \rfiveh\
 determined directly from the model and that estimated from a core-excluded
 global spectrum. We compare cool core and non-cool core clusters in terms
 of the logarithmic slopes of their gas density and temperature profiles
 and the distribution of model parameters and conclude that the two
 categories are clearly separable. In particular, we confirm the
 effectiveness of the logarithmic gradient of the gas density profile
 measured at 0.04\rfiveh\ in differentiating between the two types of
 cluster.

\end{abstract}
%%%%%%%%%%%%%%%%%%%%%%%%%%%%%%%%%%%%%%%%%%%%%%%%%%%%%%%%%%%%%%%%%%%%


%%%%%%%%%%%%%%%%%%%%%%%%%%%%%%%%%%%%%%%%%%%%%%%%%%%%%%%%%%%%%%%%%%%%
\section{Introduction}
\label{sec:intro}
Given the dominance of gravity in the Universe, the masses of large scale
structures such as clusters of galaxies are of great importance in
cosmological modelling \citep[e.g.][]{pre74} as well as in astrophysical
interpretation of their properties. The main methods for measuring cluster
masses directly are by gravitational lensing, virial analysis of the galaxy
velocity distribution and X-ray mapping of the pressure profile of the hot
intracluster medium \citep[ICM; e.g. see the review by][and references
therein]{voit05}.

Using X-ray observations to measure cluster masses has the advantage that
clusters present a sharp contrast against the X-ray background (which is
itself relatively faint and uniform on cluster angular scales), since their
(mainly bremsstrahlung) emissivity scales with $\rmsub{\rho}{gas}^2
\sqrt{T}$, and the ICM is compressed and strongly heated in clusters,
raising both these quantities by large factors. Moreover, a number of
studies based on cosmological simulations have validated the crucial
assumption of hydrostatic equilibrium in the X-ray modelling of clusters,
quantifying the systematic bias caused by non-thermal pressure support in
the range $\sim$5--20 per cent, depending on dynamical state
\citep{evrard96,rasia06,nagai07,piffaretti08,lau09}.

However, measuring X-ray cluster masses directly can be challenging and
typically requires high-quality observations, restricting the number of
objects for which reliable masses can be obtained and typically favouring
the most massive and/or luminous clusters. The usual approach is to model
the (deprojected) gas temperature, $T(r)$, and density, $\rho(r)$,
separately and combine them to infer the mass profile under the assumption
of hydrostatic equilibrium (HSE), via the following equation
\begin{equation}
M_{\mathrm{grav}}\left(r\right)=-\frac{kT\left(r\right)r}{G\mu
 \rmsub{m}{p}}\left[\frac{\mathrm{d}\ln{\rho}}{\mathrm{d}\ln{r}}+\frac{\mathrm{d
}\ln{T}}{\mathrm{d}\ln{r}}\right],
\label{eqn:M(r)}
\end{equation}
\citep[e.g.][]{fabricant80}, where $\mu$ is the mean molecular weight of the 
gas and \rmsub{m}{p} is the proton mass. Alternatively, the pressure
gradient can be estimated directly and combined with the density to
calculate the mass \citep[e.g.][]{voigt06}. The resulting values can then
be separately fitted with a mass model such as the NFW profile
\citep{navarro95}. This approach treats the density and temperature (or 
pressure) separately, and thus does not exploit the natural coupling
between these quantities that is implied by the assumption of HSE in a
well-behaved gravitational potential. Moreover, the inferred mass
distribution can have negative values, particularly if the gas profiles are
insufficiently smooth.

% latex table* generated in R 2.9.1 by xtable 1.5-5 package
% Mon Aug 17 12:19:31 2009
\begin{table*}
\begin{center}
\begin{tabular}{lcccccccc}
  \hline
Name & Mean $kT$ (keV) & $T_0$ (keV) & $t$ & $a$ (kpc) & $\alpha$ & $f$ & \rfiveh\ (kpc) & Cool-core status \\ 
  \hline
NGC 5044 & 1.17$_{-0.05}^{+0.04}$ & 3.40$\pm2.08$ & 0.17$\pm0.08$ & 404$\pm285$ & 0.37$\pm0.14$ & 0.01$\pm0.02$ & 717$\pm254$ & CC \\ 
  Abell 262 & 2.08$_{-0.09}^{+0.11}$ & 2.74$\pm0.18$ & 0.21$\pm0.12$ & 333$\pm53$ & 0.05$\pm0.02$ & 0.52$\pm0.13$ & 634$\pm27$ & CC \\ 
  Abell 1060 & 2.92$_{-0.11}^{+0.11}$ & 4.42$\pm0.10$ & 0.73$\pm0.03$ & 314$\pm19$ & 0.10$\pm0.07$ & 0.38$\pm0.04$ & 770$\pm14$ & non-CC \\ 
  Abell 4038 & 3.04$_{-0.09}^{+0.07}$ & 4.66$\pm0.08$ & 0.52$\pm0.06$ & 349$\pm14$ & 0.09$\pm0.02$ & 0.54$\pm0.05$ & 796$\pm6$ & non-CC \\ 
  Abell 1367 & 3.22$_{-0.18}^{+0.18}$ & 3.84$\pm0.30$ & 0.76$\pm0.32$ & 1200$\pm659$ & 0.03$\pm0.13$ & 2.35$\pm1.66$ & 768$\pm59$ & non-CC \\ 
  Abell 2147 & 3.69$_{-0.18}^{+0.18}$ & 5.88$\pm0.54$ & 0.71$\pm0.09$ & 1007$\pm143$ & 0.52$\pm0.72$ & 0.45$\pm0.37$ & 968$\pm44$ & non-CC \\ 
  2A 0335+096 & 4.09$_{-0.13}^{+0.13}$ & 6.53$\pm0.36$ & 0.00$\pm0.02$ & 648$\pm67$ & 0.10$\pm0.01$ & 0.47$\pm0.04$ & 998$\pm34$ & CC \\ 
  Abell 2199 & 4.50$_{-0.24}^{+0.20}$ & 7.01$\pm0.43$ & 0.16$\pm0.03$ & 605$\pm124$ & 0.08$\pm0.02$ & 0.63$\pm0.09$ & 1025$\pm47$ & CC \\ 
  Abell 496 & 4.80$_{-0.14}^{+0.15}$ & 7.26$\pm1.70$ & 0.17$\pm0.03$ & 769$\pm339$ & 0.15$\pm0.05$ & 0.32$\pm0.17$ & 1064$\pm152$ & CC \\ 
  Abell 1795 & 5.62$_{-0.35}^{+0.36}$ & 9.41$\pm0.86$ & 0.18$\pm0.04$ & 1032$\pm273$ & 0.16$\pm0.02$ & 0.50$\pm0.08$ & 1206$\pm56$ & CC \\ 
  Abell 3571 & 6.41$_{-0.23}^{+0.23}$ & 9.51$\pm0.18$ & 0.21$\pm0.21$ & 501$\pm24$ & 0.02$\pm0.01$ & 0.83$\pm0.05$ & 1134$\pm15$ & non-CC \\ 
  Abell 2256 & 6.52$_{-0.36}^{+0.39}$ & 17.27$\pm5.36$ & 0.35$\pm0.11$ & 4176$\pm1717$ & 0.75$\pm0.30$ & 0.06$\pm0.03$ & 1433$\pm100$ & non-CC \\ 
  Abell 85 & 6.64$_{-0.20}^{+0.20}$ & 8.94$\pm0.58$ & 0.14$\pm0.03$ & 825$\pm121$ & 0.08$\pm0.02$ & 0.85$\pm0.15$ & 1164$\pm41$ & CC \\ 
  Abell 3558 & 7.17$_{-0.46}^{+0.49}$ & 9.60$\pm0.65$ & 0.45$\pm0.14$ & 879$\pm157$ & 0.11$\pm0.06$ & 0.72$\pm0.26$ & 1214$\pm44$ & non-CC \\ 
  Abell 3667 & 7.60$_{-0.37}^{+0.38}$ & 11.88$\pm0.66$ & 0.40$\pm0.07$ & 2374$\pm339$ & 0.11$\pm0.05$ & 1.38$\pm0.54$ & 1305$\pm54$ & non-CC \\ 
  Abell 478 & 8.23$_{-0.26}^{+0.26}$ & 12.57$\pm0.76$ & 0.14$\pm0.03$ & 886$\pm174$ & 0.12$\pm0.01$ & 0.74$\pm0.09$ & 1351$\pm49$ & CC \\ 
  Abell 3266 & 8.38$_{-0.43}^{+0.67}$ & 11.56$\pm0.56$ & 0.00$\pm0.21$ & 1290$\pm135$ & 0.02$\pm0.01$ & 1.23$\pm0.15$ & 1345$\pm32$ & non-CC \\ 
  Abell 2029 & 8.96$_{-0.30}^{+0.30}$ & 12.29$\pm1.49$ & 0.21$\pm0.02$ & 785$\pm207$ & 0.13$\pm0.03$ & 0.60$\pm0.13$ & 1327$\pm104$ & CC \\ 
  Abell 401 & 9.16$_{-1.06}^{+1.41}$ & 12.43$\pm0.86$ & 0.66$\pm0.23$ & 964$\pm190$ & 0.15$\pm0.14$ & 0.94$\pm0.17$ & 1362$\pm63$ & non-CC \\ 
  Abell 2142 & 9.50$_{-0.42}^{+0.43}$ & 16.97$\pm1.14$ & 0.35$\pm0.03$ & 1248$\pm164$ & 0.21$\pm0.04$ & 0.50$\pm0.11$ & 1591$\pm63$ & non-CC \\ 
   \hline
\end{tabular}
\caption{Best-fit model parameters for each cluster and
corresponding \rfiveh, listed in order of increasing mean temperature as
measured in \citetalias{san09}. Errors are 1$\sigma$, estimated
from 200 bootstrap resamples. The cool core status is as determined 
in \citet{san06}.}
\label{tab:main}
\end{center}
\end{table*}
		% tab:main

Another method is to use a model for the total mass density and either a
parametric \citep[e.g.][]{lloyd-davies00} or non-parametric
\citep[e.g.][]{allen98a,schmidt07} form for the gas density to 
predict $T(r)$ (assuming HSE), which is then fitted to the observed data.
However, this method can yield unphysical behaviour in the form of negative
temperatures.

Recently, \citet[hereafter \citetalias{ascasibar08}]{ascasibar08} have
developed a simple, phenomenological cluster model, based on the
\citet{hernquist90} mass profile, which avoids unphysical behaviour 
and provides a full description of the gas and mass distributions with only
five adjustable parameters. The model can be simultaneously fitted to both
the gas temperature and density profiles, providing additional stability
and enabling it to be applied to even relatively poor quality data. In this
paper we assess the performance of this model as applied to a statistical
sample of clusters -- that of \citet{san06} and \citet[][hereafter
\citetalias{san09}]{san09}, which incorporates both cool core (CC) and 
non-CC clusters. We investigate the effectiveness of the model through
analysis of the residuals from the best-fit and explore some of its
capabilities in measuring derived cluster properties.

Throughout this paper we adopt the following cosmological parameters:
$H_{0}=70$\kmpspMpc, $\rmsub{\Omega}{m}=0.3$ and
$\Omega_{\Lambda}=0.7$. All errors are 1$\sigma$, unless otherwise stated.

%%%%%%%%%%%%%%%%%%%%%%%%%%%%%%%%%%%%%%%%%%%%%%%%%%%%%%%%%%%%%%%%%%%%
\section{Cluster sample}
The objects studied in this paper comprise the statistical sample of 20
galaxy clusters observed with \Chandra\ presented in \citet{san06} and
\citetalias{san09}, where details of the selection and the basic properties 
of this sample can be found.  Briefly, it consists of 20 clusters drawn
from the flux-limited catalogue of \citet{ikebe02}, excluding the Coma,
Fornax and Centaurus clusters, owing to their very large angular sizes.  It
is interesting to note that, while half of the sample show some indication
of merging activity \citep{san06}, there are no seriously disturbed
clusters of the type that would prevent a 1-dimensional radial analysis,
unlike the 2 (out of 33) highly irregular clusters present in the
representative \XMM\ cluster structure survey (REXCESS), for example
\citep{boehringer07,croston08}.

%--
\subsection{Spectral profiles and errors}
\label{sec:spec_profile}
Deprojected gas temperature and density profiles were obtained in discrete
annular bins as described in \citetalias{san09}, but an improved method was
used here to evaluate errors on these quantities, given the large
interdependence between radial bins in the \textsc{projct} scheme in
\textsc{xspec}. A series of 200 Poisson realizations of the best-fitting 
\textsc{projct} model were generated, using the same background and 
response files, to produce a series of simulated source spectra. Each set
of annular spectra was then fitted with the \textsc{projct} model and
treated like the original data, using the same corresponding background
spectra and response files. This produced a suite of simulated measurements
of gas temperature and density in each annulus, which were used to evaluate
the error in each radial bin using the median absolute deviation (MAD) of
the values, which is a robust estimator of the standard deviation. The MAD
is particularly well suited to heavy-tailed distributions, although it is
less efficient when applied to Gaussian ones \citep[e.g. see][for a
discussion]{beers90}.

\begin{figure*}
  \centering \subfigure{
  \includegraphics[width=11cm]{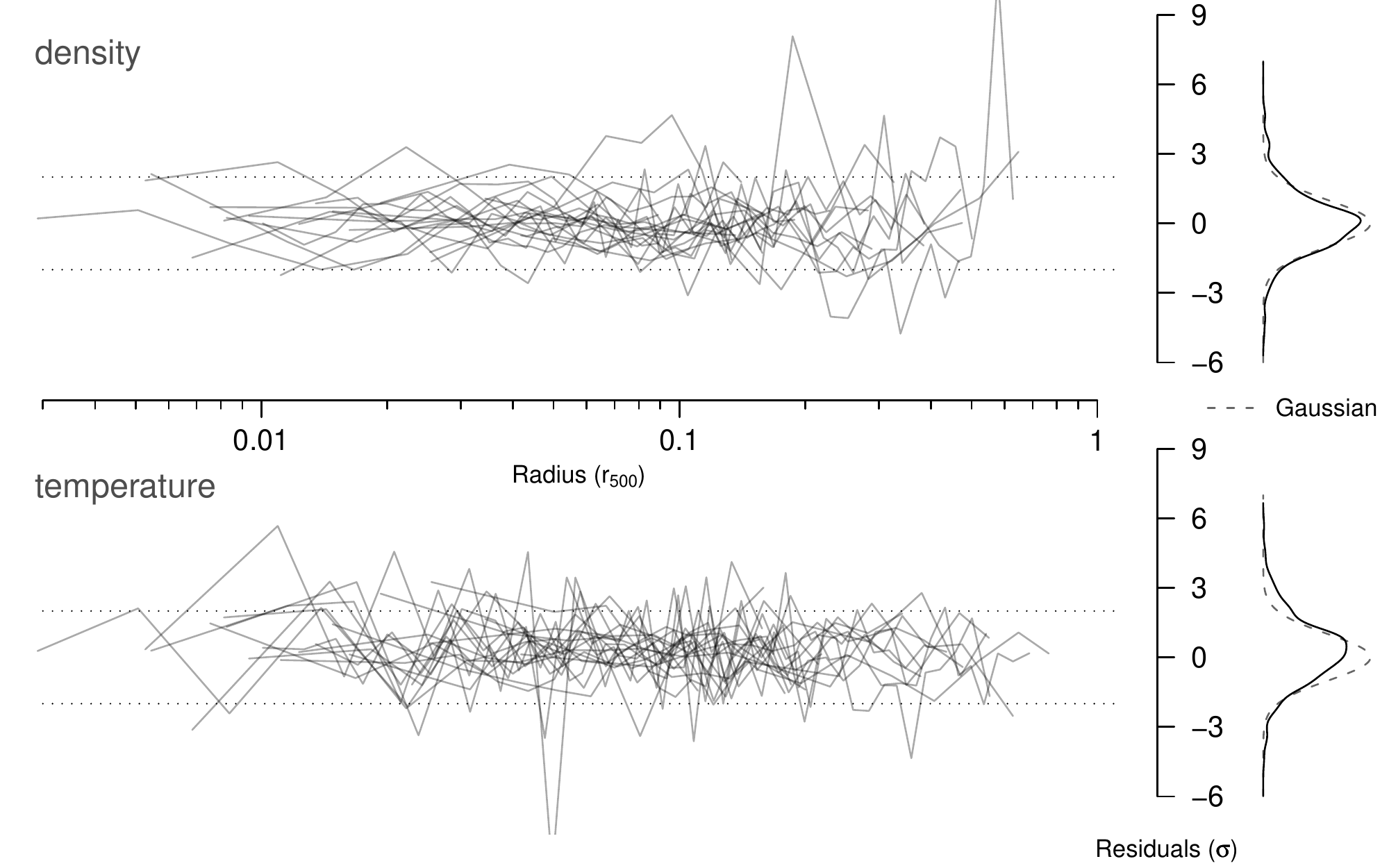}} \hspace{0cm}
  \subfigure{
  \includegraphics[width=6.3cm]{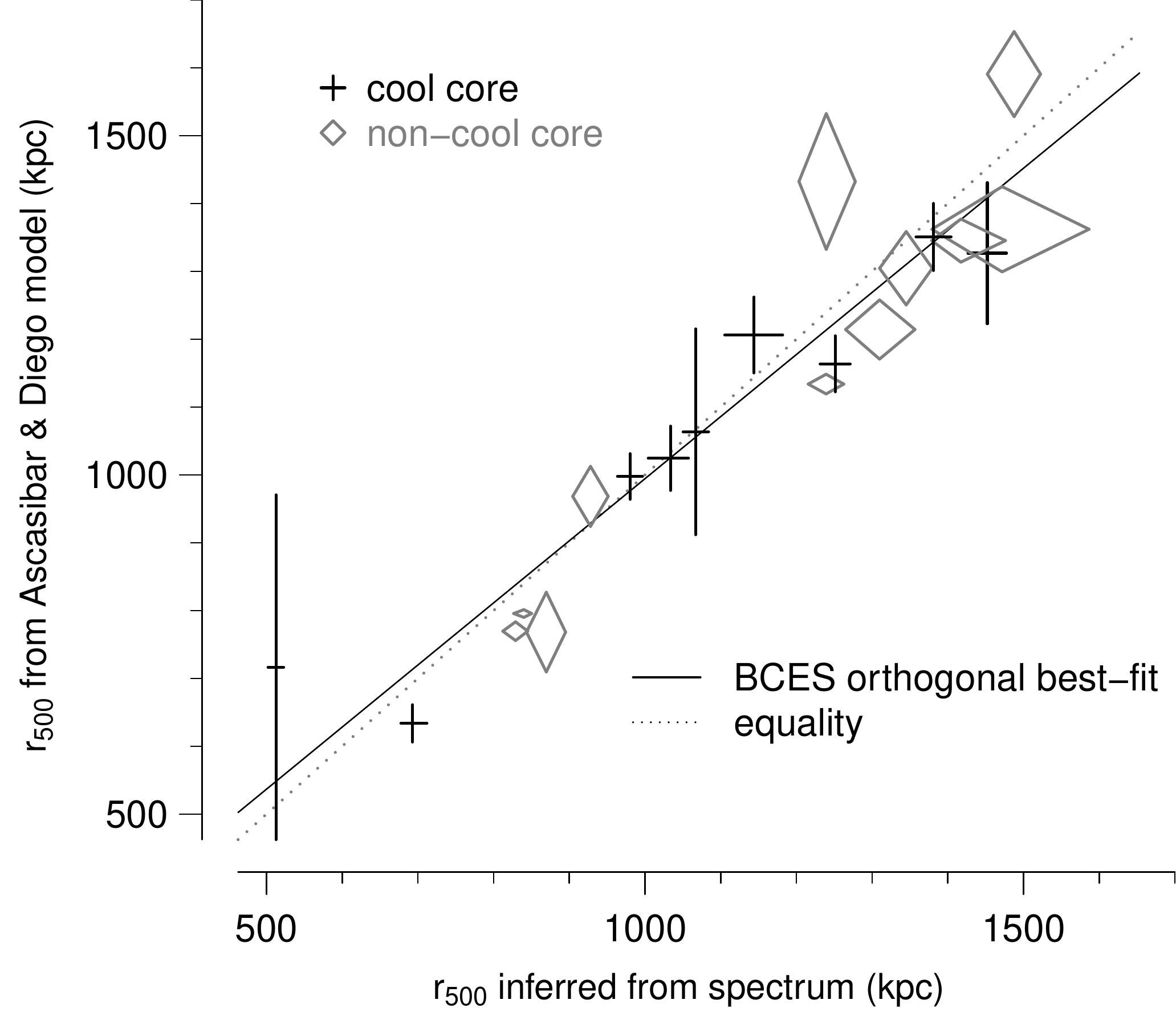}}
  \caption{\textit{Left:} A comparison of the residuals from the
  best-fitting \citetalias{ascasibar08} model, normalized by the
  measurement errors on each point, as a function of scaled radius for both
  the gas density and temperature data. Each line represents a different
  cluster and shows the fluctuations about the model purely in terms of the
  statistical errors on the observed data points. Dotted lines mark
  $\pm2\sigma$, and the marginal distribution of the data is depicted as a
  kernel smoothed density estimate, compared to a Gaussian of unit
  variance. \textit{Right:} The comparison between \rfiveh\ calculated from
  the cluster model and from the mean temperature estimated in
  \citetalias{san09}. The solid line is the best fit BCES weighted
  orthogonal regression, with a slope of $0.91\pm0.11$ and the dotted line
  is the locus of equality.}  \label{fig:resid-compare_r500}
\end{figure*}

%%%%%%%%%%%%%%%%%%%%%%%%%%%%%%%%%%%%%%%%%%%%%%%%%%%%%%%%%%%%%%%%%%%%
\section{Cluster model}
\label{sec:model}
We use the phenomenological cluster model of \citetalias{ascasibar08},
which describes the gas density and temperature profile in a state of
hydrostatic equilibrium in a \citet{hernquist90} gravitational potential
with just five free parameters.  The model assumes a polytropic ICM, with
central gas temperature, $T_0$, modified by a variable cool core component
controlled by the parameter $t$ ($0<t<1$), which becomes important inside a
radius a fraction $\alpha$ ($0<\alpha<1$) times the dark matter scale
radius, $a$. The gas density normalization is expressed in terms of a
fraction, $f$, of the cosmic mean baryon fraction. The model yields
smoothly varying profiles which avoid unphysical behaviour, such as
negative values, while capturing the typically observed cluster properties
with good accuracy-- e.g. the single-peaked gas temperature profile
commonly found in cool-core clusters.

The use of a \citet{hernquist90} parametrization for the mass profile has
the advantage of yielding a convergent projected distribution (since the
mass density falls off at large radii as $r^{-4}$), unlike the
\citet{navarro95} profile, for example. This property enables direct 
comparison of projected X-ray masses derived from the model with those
measured using gravitational lensing within a fixed aperture. Such a
comparison is presented in \citet{richard09}, based on the
\citetalias{ascasibar08} model analysis of \citet{san09b}, and reveals
generally good agreement between the X-ray and strong gravitational lensing
projected masses within a cylinder of radius 250\,kpc, particularly for
clusters without evidence of significant dynamical disturbance.

%--
\subsection{Model fitting}
The cluster model is fitted simultaneously to the observed temperature and
density profiles and the $\chisq$ statistic is evaluated and minimized for
both datasets combined. This joint fit introduces extra stability into the
minimization and provides tighter constraints on the model parameters,
exploiting the coupling between the gas density and temperature implied by
the assumption of HSE.

Asymmetric errors on $T(r)$ and $\rhogas(r)$ are handled by using the upper
measurement error when the model lies above the measured value and the
lower measurement error otherwise. Since the model is not expected to
perform well on small scales \citepalias[e.g. see][]{ascasibar08}, where
the effects of stellar baryonic mass become non-negligible, data points
within 5\,kpc are excluded from the fit. In the case of 2A0335+096 and
Abell~262, larger exclusion radii were used, of 20 and 8\,kpc respectively,
within which a clear deviation of the data from the model was evident,
associated with a departure from HSE as a result of disruptions in the
cores of these clusters \citep{mazzotta03,clarke09}. The best-fit model
parameters and 1$\sigma$ errors are listed in Table~\ref{tab:main}.

%--
\subsection{Error estimates}
One of the main challenges in modelling cluster properties is the process
of deriving error estimates, on both the parameters of the model itself and
on any derived quantity evaluated using those parameters. Errors are
particularly important when fitting scaling relations with weighted
regression techniques and when attempting to determine intrinsic scatter in
the measurement of some quantity over and above the purely statistical
dispersion.  Although there are many reliable ways of determining error
estimates for fitted model parameters, these do not provide errors directly
on \emph{derived} quantities, which are usually desirable in a cluster
modelling analysis (e.g. measuring \rfiveh, or the gas fraction within this
radius).

In this situation, a common approach is to generate a set of Monte Carlo
(MC) realizations of the original data, which can be used to produce a
corresponding set of best-fit models. Errors on any parameter of the model
or derived quantity are then determined from the spread in values obtained,
typically measured using the standard deviation.  However, the success of
this method depends critically on the prescription used to generate the MC
realizations, which must produce realistic mock datasets that capture the
essential properties of the original measurements.

\begin{figure*}
  \centering \subfigure{
  \includegraphics[width=8.7cm]{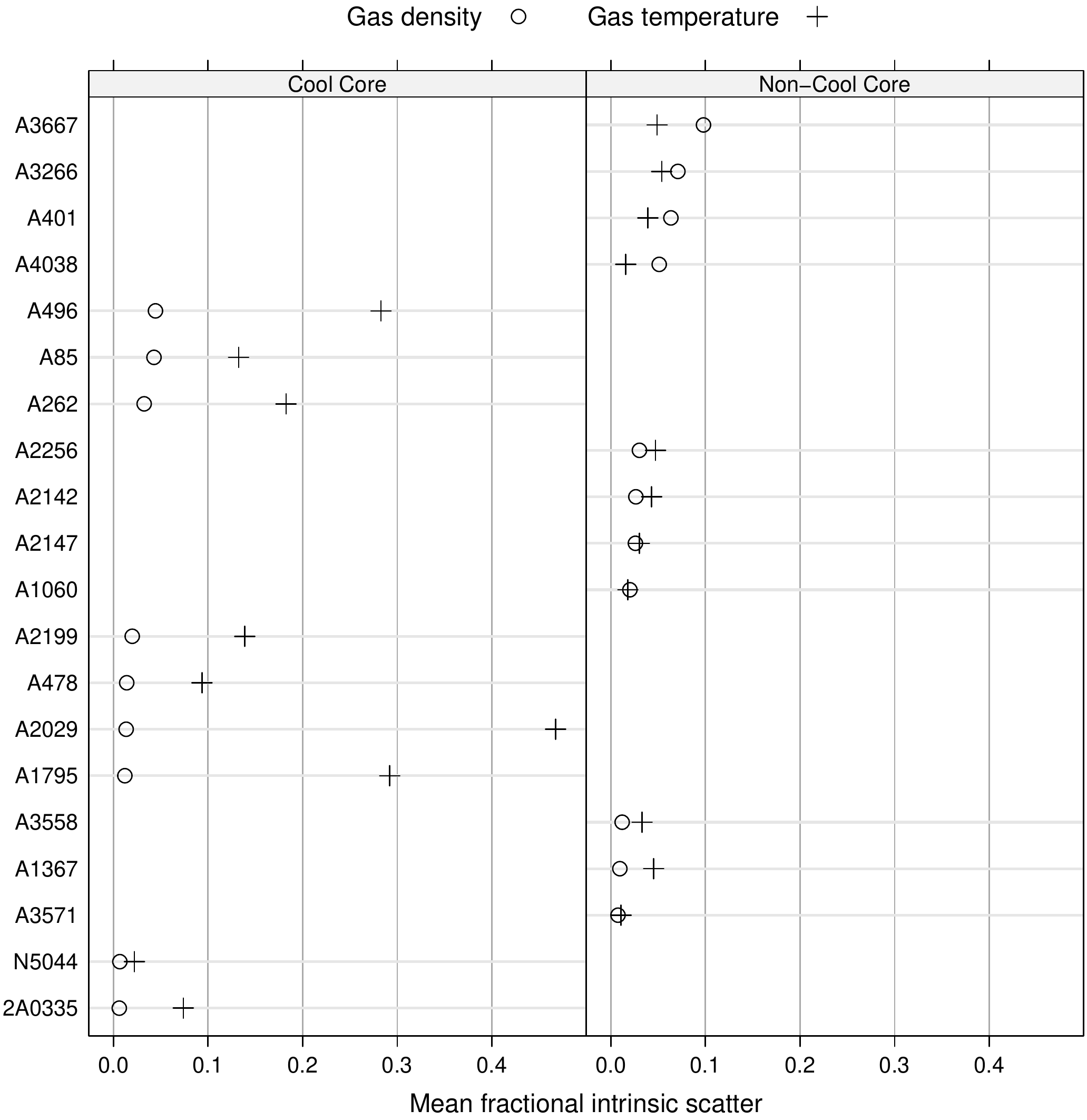}} \hspace{0cm}
  \subfigure{
  \includegraphics[width=8.7cm]{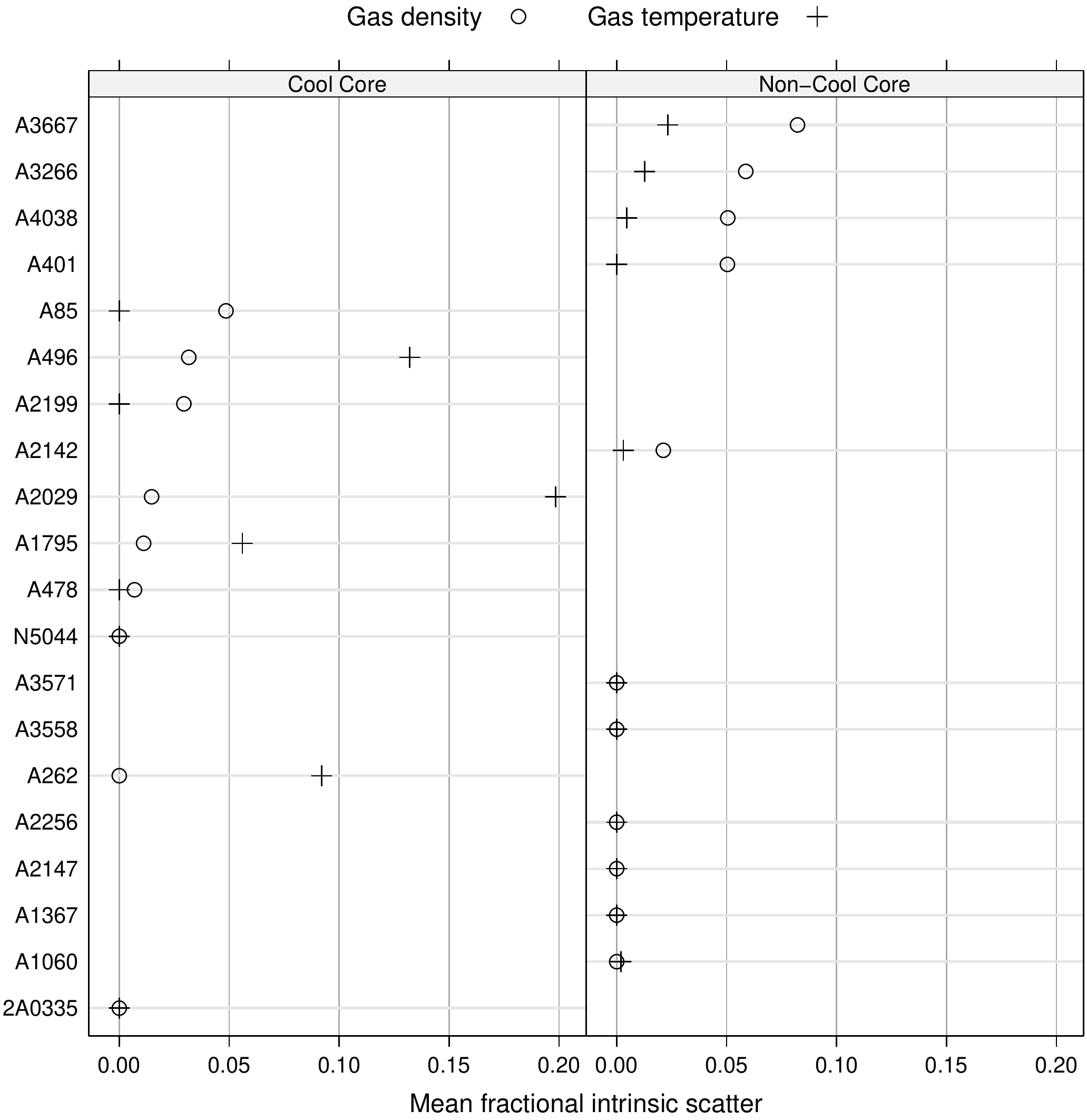}}
  \caption{Dot plots of the mean fractional intrinsic scatter about the
  best-fit model for each cluster, in both the gas density and temperature
  for the original (\textit{left} panel) and rebinned (\textit{right}
  panel) profiles. In each panel the clusters are arranged in order of
  decreasing scatter in density from top to bottom and are separated
  according to their CC status.} \label{fig:dotplot_scatterfrac}
\end{figure*}

Often what is done is simply to perturb the observed data, using the
measured value as the mean for a Gaussian random number generator, with the
measurement error as the standard deviation. However, this method has two
drawbacks. First of all, the measured values already include some random
noise and thus are imperfect estimates of the expectation value for the
distribution at the radius of the measurement. For example, in the case of
an extreme outlier, all MC realizations based on this value would
automatically `start' at an extreme point compared to the underlying
distribution that generated the data point (and compared to which it was
judged to be an outlier), thus biasing them significantly. This problem
could be mitigated by using the predicted value from the best-fit model as
the mean for the Gaussian random perturbation.

The second problem with the above approach is that it assumes that the data
points are Gaussian-distributed about the best-fit with a dispersion given
by the measurement errors alone. No allowance is made for intrinsic scatter
in the data, which is a common situation in astronomical analyses. For
example, in the case of modelling galaxy clusters, the frequent assumption
of spherical or even ellipsoidal geometry will always be violated at some
level as a result of substructures and other features, which add scatter to
azimuthal profiles in addition to that from measurement
uncertainties. Therefore, to avoid both these problems we use bootstrap
resampling to generate a set of realizations of the observed data.

\subsection{Bootstrap resampling}
A set of $N$ realizations of the original data were generated by separate
bootstrap resampling of the temperature and density data, with $N=200$. The
cluster model was then fitted to each realization dataset, treating it
identically to the original data, to produce a series of $N$ values of the
model parameters. In calculating derived quantities, such as \rfiveh, the
quantity was evaluated separately for each bootstrap realization, to yield
$N$ values. The uncertainty on any given quantity was then obtained from
the spread in the bootstrapped values, as measured using the median
absolute deviation, as described in Section~\ref{sec:spec_profile} above.
In a few cases, where the best-fit and many of the bootstrap values were at
the lower limit of the parameter (e.g. $t=0$ for 2A~0335+096
\& Abell~3266; see Table~\ref{tab:main}), the MAD produced unrealistically
low errors. In these cases we calculated the error using the interquartile
range of the $N$ bootstrap realizations, adjusted for a normally consistent
estimate of the standard deviation by dividing by $2\times$ qnorm(3/4)
($\approx1.35$), where qnorm is the normal quantile distribution. This was
done wherever the MAD was less than half the adjusted IQR value.

%%%%%%%%%%%%%%%%%%%%%%%%%%%%%%%%%%%%%%%%%%%%%%%%%%%%%%%%%%%%%%%%%%%%
\section{Testing the model}
\label{sec:model_test}
The left panel of Fig.~\ref{fig:resid-compare_r500} shows the residuals
from the model for both the gas temperature and density, as a function of
scaled radius. There is no indication of any significant systematic trends
with radius and the marginal distribution of the residuals is consistent
with the Gaussian distribution expected from the measurement errors alone,
apart from a slight excess of positive temperature residuals \citep[as also
seen in][]{san09b}. Note that the gas density profiles extend only to the
penultimate spectral bin, owing to the non-trivial volume element
associated with the outermost annulus in the deprojection \citep{san06}.

To assess the reliability of the cluster model, we have compared the
predicted \rfiveh\ values for each cluster with those obtained from an
independent method in \citetalias{san09}, based on a single spectral
fit. Briefly, the \rfiveh\ values from \citetalias{san09} were determined
iteratively by extracting a spectrum in the range 0.15--0.2\rfiveh, fitting
an \textsc{apec} hot plasma model to it and determining \rfiveh\ from the
resulting temperature, based on the \MT\ relation of \citet{vikhlinin06},
adjusted for the effect of using a different aperture. The right panel of
Fig.~\ref{fig:resid-compare_r500} shows the comparison of \rfiveh\ obtained
with the two methods and it is clear that there is generally very good
agreement. A weighted orthogonal linear regression using the BCES method of
\citet{akritas96} yields a best-fit slope of $0.91\pm0.11$, using a pivot
point set at the mean of all the $X$ and $Y$ values, so as to decouple the
measurement errors on the slope and intercept as far as possible. The
corresponding best-fit intercept is $-15\pm20$, which is fully consistent
with a zero offset between the two methods.

%--
\subsection{Residual real variance about the best-fit}
Although the model appears to be a good representation of the data,
departures from the fit are expected, since real clusters are not perfectly
spherical or fully in HSE. Such deviations can be quantified in terms of
the real variance of the data about the model, beyond that expected from
the measurement errors, i.e. the intrinsic scatter. As described in
Appendix~\ref{sec:appendix}, we calculate the mean fractional intrinsic
scatter about the best fitting gas temperature and density profiles using
Equations~\ref{eqn:s} \& \ref{eqn:f.mean}. The left panel of
Fig.~\ref{fig:dotplot_scatterfrac} shows the mean values of the intrinsic
scatter in gas density and temperature as a dot plot, with the clusters
listed in order of decreasing scatter in density from top to bottom, which
is below 10 per cent in all cases. The median values across the sample are
0.023 and 0.048 for the density and temperature, respectively.

\begin{figure}
\centering
\includegraphics[width=8.4cm]{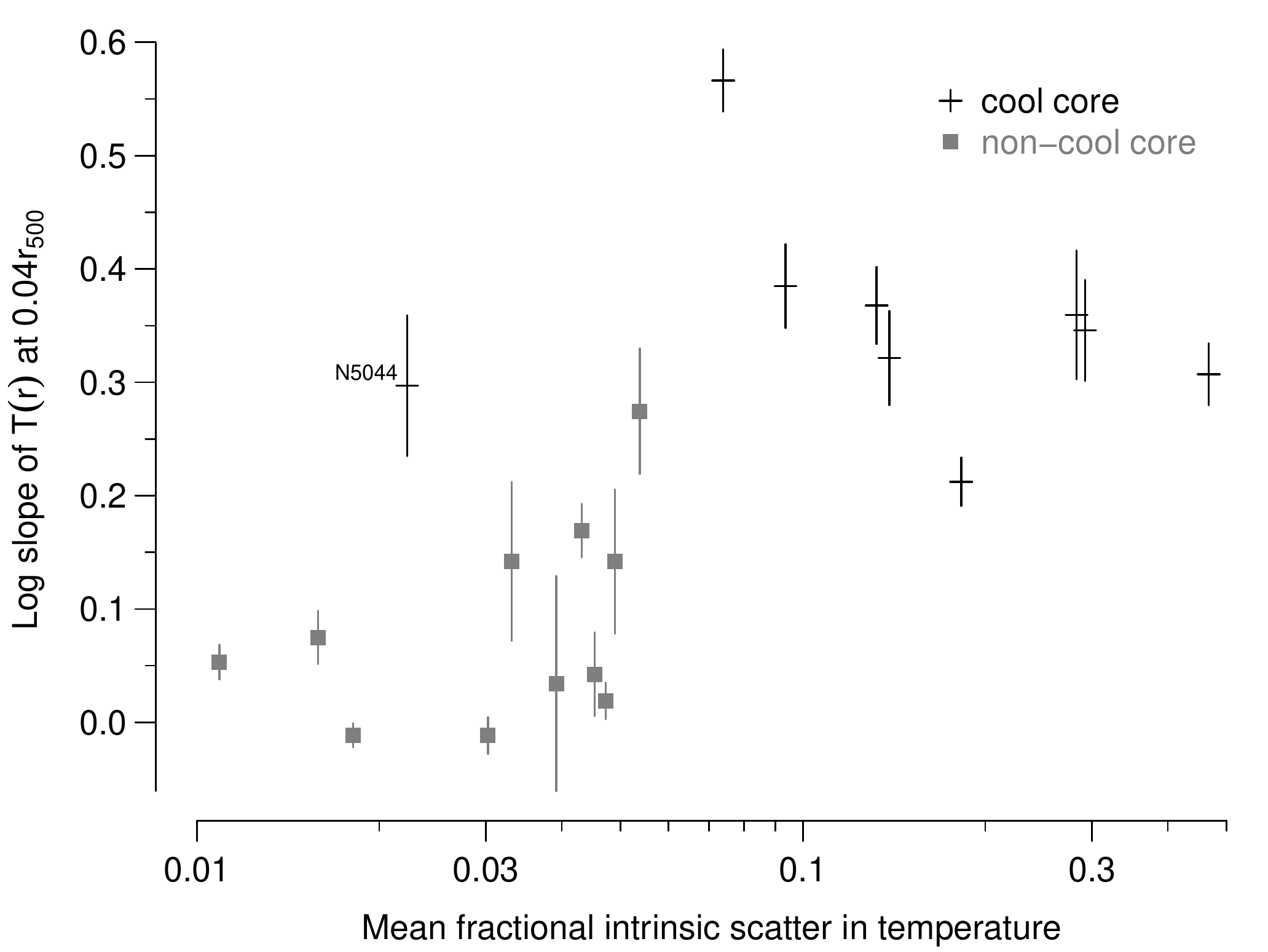}
\caption{ \label{fig:logTgrad-kTscatter}
The relationship between the mean fractional intrinsic scatter from the
model in temperature and the logarithmic slope of the temperature profile
at 0.04\rfiveh, with the outlier galaxy group NGC~5044 labelled (see text
for details).}
\end{figure}

It is striking from Fig.~\ref{fig:dotplot_scatterfrac} that the non-CC
clusters have a broadly similar mean fractional intrinsic scatter in both
temperature and density, whereas most of the CC clusters have quite high
scatter in temperature. This is caused by instabilities in the
deprojection, possibly as a result of non-spherical geometry or multiphase
gas \citep{fabian06a}, which result in oscillations in the temperature
profile recovered using the \textsc{xspec} \textsc{projct} model
\citep{russell08}.  Nevertheless, it is clear from
Fig.~\ref{fig:resid-compare_r500} that any such fluctuations are small
\emph{compared to the measurement errors} and that the model performs well in
describing the data. Furthermore, this noise can easily be smoothed to
clearly reveal the underlying trend \citep{san06}.

The impact of oscillatory fluctuations on the intrinsic scatter can be
gauged by rebinning the original profiles, grouping pairs of bins together
and recalculating the mean fractional intrinsic scatter about the original
best fitting profiles. The error in each new bin is calculated by adding in
quadrature the errors on the constituent bins and a characteristic radius
is assigned based on the inner- and outermost bounding radii of the
constituent bins, as per \citet{san06}. This process will suppress
oscillatory behaviour, which causes neighboring bins to be anticorrelated,
but will not significantly alter any other systematic deviations. The
corresponding dot plot for the rebinned profiles is plotted in the right
panel of Fig.~\ref{fig:dotplot_scatterfrac} and it can be seen that the
fractional scatter in the temperature is reduced markedly for the CC
clusters, while the density scatter is mostly unaffected. In particular,
the cluster ranking by density scatter is largely preserved. This
demonstrates that the intrinsic scatter in density is more representative
of bulk systematic deviations from the model, whereas oscillatory
deprojection instabilities can contribute significantly to the temperature
scatter, particularly for CC clusters.

Such instabilities in the deprojection are mainly confined to the cool core
itself, where the temperature gradient is steep \citep[e.g.][]{san06}.
This can be seen in Fig.~\ref{fig:logTgrad-kTscatter}, which shows a
significant correlation between the scatter in temperature (from the
original profiles) and the logarithmic slope of the gas temperature profile
at 0.04\rfiveh\ (see Section~\ref{sec:gas_gradients}), demonstrating that
the clusters with the largest intrinsic scatter all have steep core
temperature gradients. The only exception is also the only galaxy group in
the sample (NCG~5044). Such cool objects radiate predominantly line
emission which provides much tighter constraints on the temperature than
the slope of the bremsstrahlung continuum, since line ratios are very
sensitive to temperature. This property enables a cleaner separation of the
different temperature components projected onto inner annuli from hotter
outer shells in cool cores dominated by line emission (i.e. $kT\la2$\,keV),
and results in smoother profiles and less intrinsic scatter about the
best-fit model.

Notwithstanding the impact of deprojection oscillations on the temperature
profile, it is apparent that the mean fractional intrinsic scatter in the
gas \emph{density} serves to quantify the extent to which the model
provides a good description of the data. For example, three of the four
clusters with mean fractional intrinsic scatter in the density exceeding 5
per cent are known merging systems: Abell~3667 \citep[e.g.][]{owers09},
Abell~3266 \citep[e.g.][]{finoguenov06} and Abell~401
\citep[e.g.][]{sakelliou04}. Such objects are known to have somewhat 
distorted morphologies and may not be in HSE. Nevertheless, even
Abell~3667-- the cluster with the largest intrinsic scatter-- has a
reasonably regular morphology that justifies the assumption of azimuthal
symmetry \citep[see for example the X-ray/optical overlay image in
figure~12 of][]{owers09}.  Furthermore, all but 3 of our sample
(Abell~2147, 2A~0335+096 and NGC~5044) were analysed by \citet{mohr99}, who
used either single or double $\beta$-model fits to ROSAT images in order to
estimate gas and total masses, finding a reasonable match to the data in
all cases.

\begin{figure}
\centering
\includegraphics[width=8.4cm]{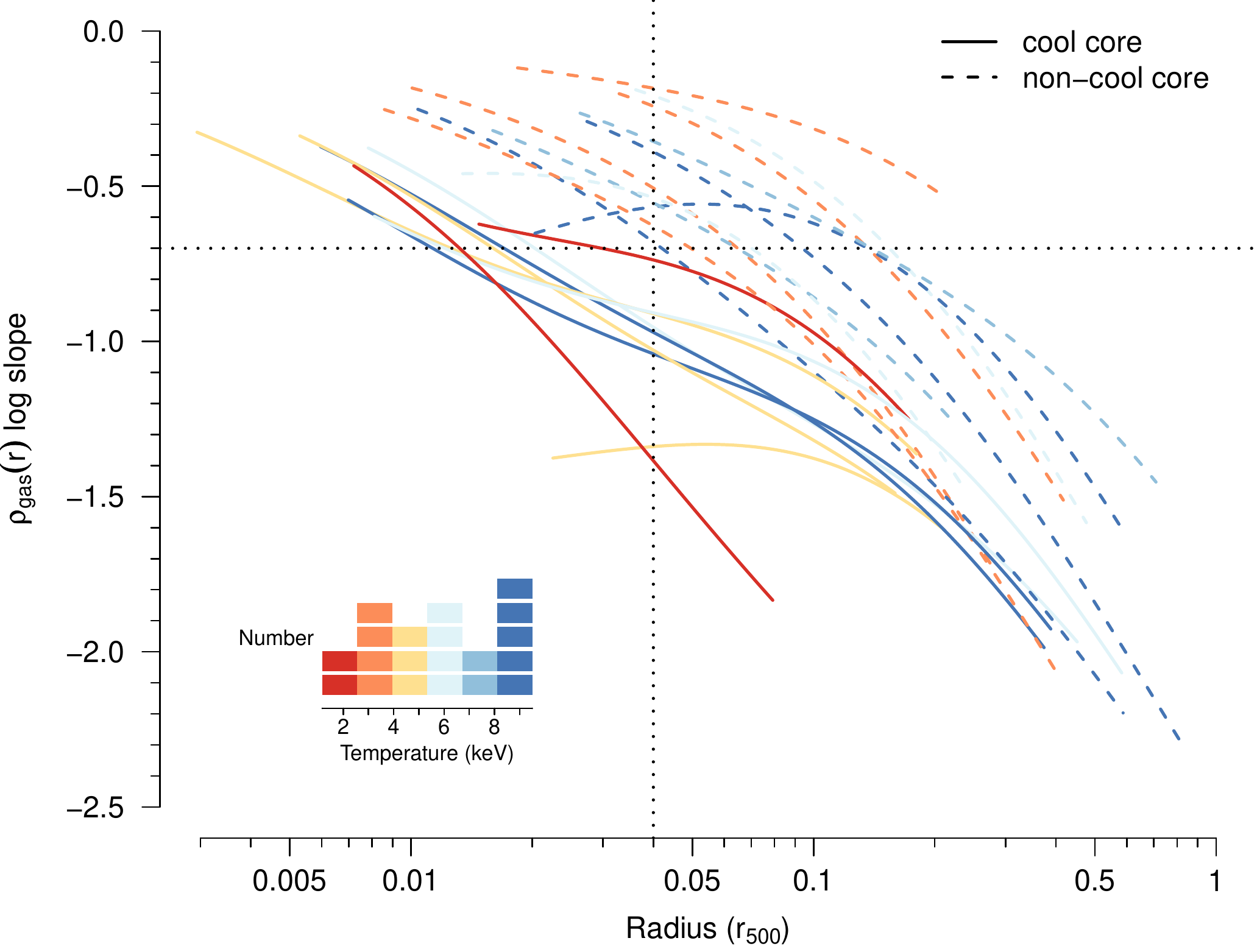}
\caption{ \label{fig:rhogasgrad(r500)}
Logarithmic slope of the gas density profile as a function of scaled
radius, coloured according to the mean cluster temperature, depicted by the
inset histogram. Curves are plotted within the radial range used to fit the
model. The dotted lines indicate a value of -0.7 at 0.04\rfiveh, which
corresponds to the value of this parameter which \citet{vikhlinin07} used
to identify strong cooling flow clusters
\citep[see also][]{san09b}.}
\end{figure}

%%%%%%%%%%%%%%%%%%%%%%%%%%%%%%%%%%%%%%%%%%%%%%%%%%%%%%%%%%%%%%%%%%%%
\section{Results}

%++
\subsection{Gas profile gradients}
\label{sec:gas_gradients}
A particularly attractive feature of well-behaved and smoothly-varying
analytic models, such as that of \citetalias{ascasibar08}, is the ability
to evaluate the gradient of predicted quantities at arbitrary radii. For
example, the gradients of the temperature and density profiles are
particularly sensitive indicators of central cooling, which causes a
progressive steepening in both cases
\citep[e.g.][]{ettori08}. Fig.~\ref{fig:rhogasgrad(r500)} shows profiles of
the gas density gradient as a function of scaled radius for the model
fits. It can be seen that inside 0.05\rfiveh\ the CC log gradients are
steeper (i.e. more negative) than those of the non-CC clusters, and in the
case of 2A~0335+096 appear consistent with a pure power law (i.e. constant
log gradient); otherwise all clusters show gradual flattening of the
density profile (log gradient profiles approaching zero).

Towards the outskirts ($\ga$\,0.3\,\rfiveh), $\rhogas(r)$ steepens
progressively with radius compared to a standard isothermal $\beta$-model,
as originally discovered by \citet{vikhlinin99}. The shape and
normalization of the profiles in Fig.~\ref{fig:rhogasgrad(r500)} are
consistent with the recent findings of \citet{croston08}, based on \XMM\
observations of a representative cluster sample, but are able to reach
smaller radii, owing to the better spatial resolution of \Chandra.

The bifurcation between CC and non-CC clusters in
Fig.~\ref{fig:rhogasgrad(r500)} motivates the use of the surface brightness
`cuspiness' as an effective proxy for cool-core status
\citep{vikhlinin07,ettori08,san09b}. In particular,
\citeauthor{vikhlinin07} identified the logarithmic gradient of the gas
density profile at 0.04\rfiveh\ as a proxy for the strength of cooling
\citep[see also][]{san09b}, which is indicated by the vertical dotted line
in Fig.~\ref{fig:rhogasgrad(r500)}: a value of $-0.7$ cleanly separates the
two categories of cluster.

\begin{figure}
\centering
\includegraphics[width=8.4cm]{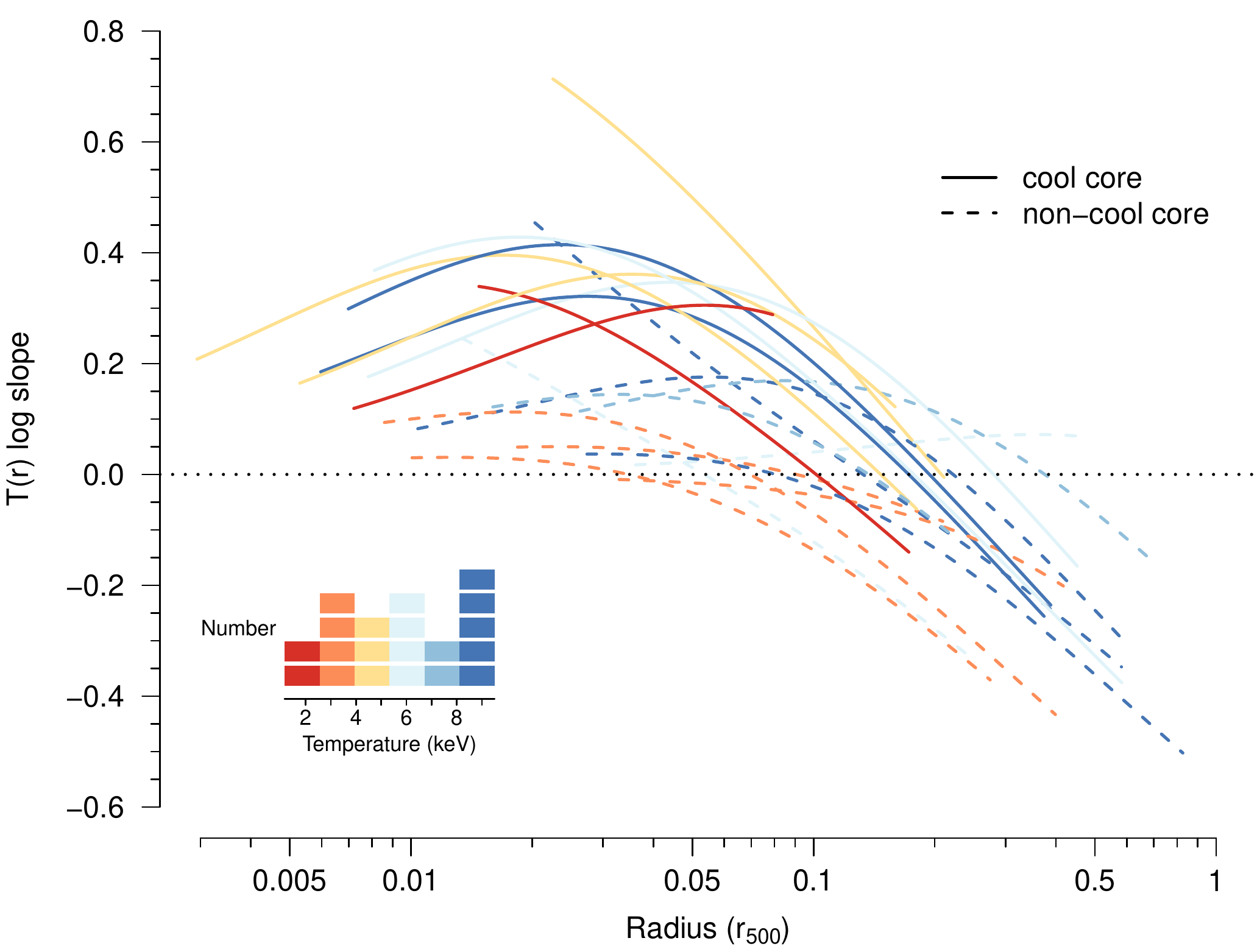}
\caption{ \label{fig:Tgrad(r500)}
Logarithmic slope of the gas temperature profile as a function of scaled
radius, coloured according to the mean cluster temperature, depicted by the
inset histogram. Curves are plotted within the radial range used to fit the
model. Intersection of the curves with the dotted zero line identifies the
radius of the peak temperature.}
\end{figure}

\begin{figure}
\centering
\includegraphics[width=8.4cm]{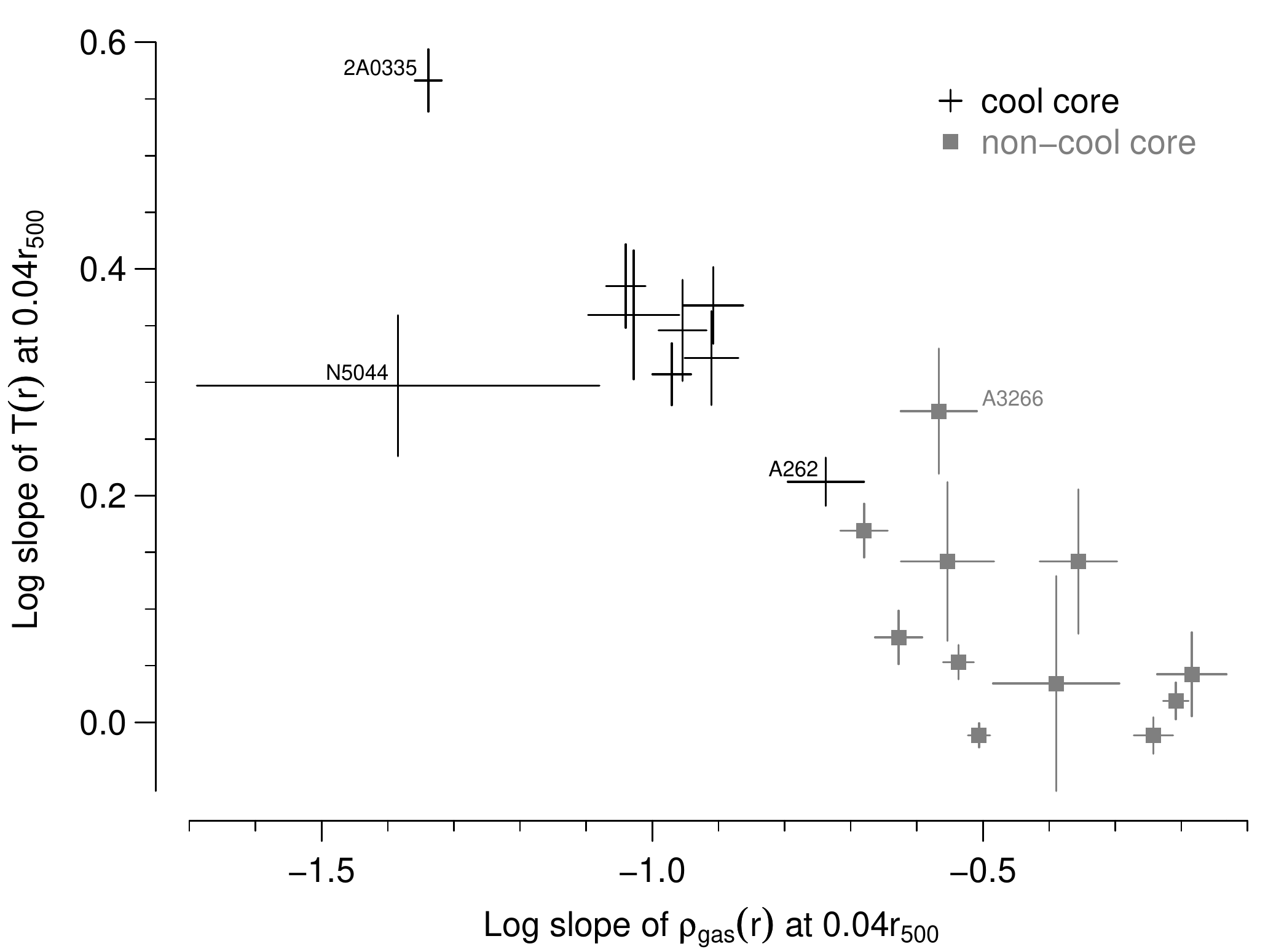}
\caption{ \label{fig:Tgrad_vs_rhogasgrad}
The relationship between the logarithmic slopes of the gas temperature and
density profiles evaluated at 0.04\rfiveh, with obvious outliers
identified.}
\end{figure}

The equivalent plot of the logarithmic gradient of the gas temperature
profile is shown in Fig.~\ref{fig:Tgrad(r500)}. Once again it can be seen
that, within 0.05\rfiveh, the CC clusters all have steeper (in this case,
positive) gradients-- the only exception is the merging, non-CC cluster
Abell~3266 (the only dashed line in Fig.~\ref{fig:Tgrad(r500)} exceeding a
logarithmic slope of 0.3). Choosing 0.04\rfiveh\ as a fiducial radius, the
relationship between the logarithmic gradient of the gas temperature and
density profiles is plotted in Fig.~\ref{fig:Tgrad_vs_rhogasgrad}, clearly
revealing the separation between the CC and non-CC clusters
\citepalias[cf.][]{san09}. The most obvious outlier is the CC poor cluster 
Abell~262, which appears to lie closer to the non-CC points. This cluster
unquestionably hosts a cool core \citep[see the temperature profile in][for
example]{san06}, but its inner core has been disrupted by repeated active 
galactic nuclei outbursts occurring on the scale of 0.04\rfiveh\ (25\,kpc)
\citep[e.g.][]{clarke09}, which could flatten the temperature and density
gradient as observed.

It is apparent from Fig.~\ref{fig:Tgrad(r500)} that all the non-CC curves
intersect the zero line, implying that their temperature profiles turn over
at least slightly at small radii; indeed, this can be seen from the
projected temperature profiles plotted in figure 3 of
\citet{san06}, as well as the fact that the values of the \citetalias{ascasibar08} 
normalized central temperature parameter, $t$, lie below unity (which would
corresd to a pure polytropic profile) in all cases
(Table~\ref{tab:main}). Nevertheless, with the exception of Abell~3266
(where the impact of merger-induced deviations from HSE may be affecting
the model fit), the effect is small and does not manifest itself in the
logarithmic slopes of the gas density profile
(Fig.~\ref{fig:rhogasgrad(r500)}).

%--
\subsection{Cool core vs. non-cool core clusters}
\begin{figure}
\centering
\includegraphics[width=8.4cm]{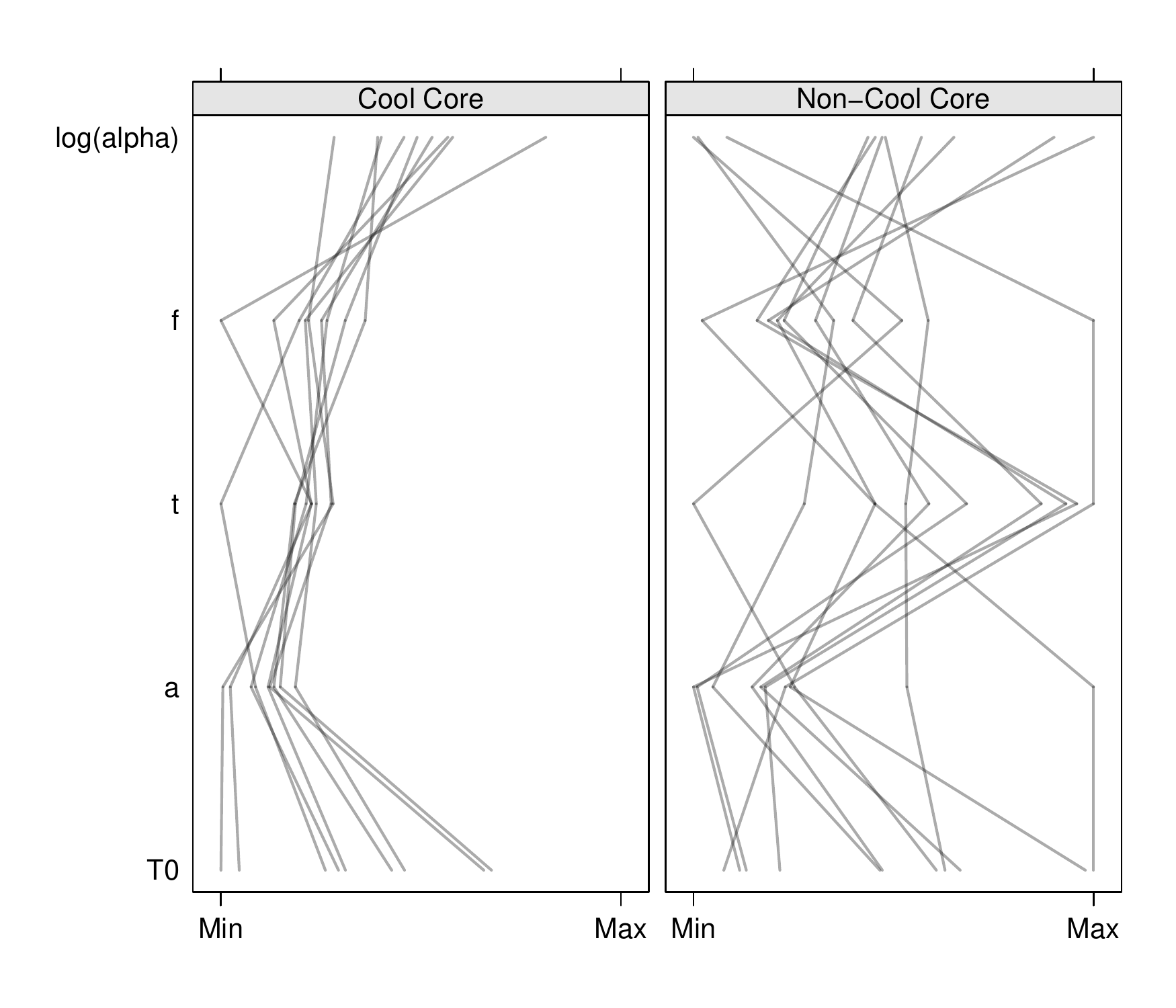}
\caption{ \label{fig:jfpars_parallel}
A parallel coordinates plot of the cluster model parameters, comparing CC
and non-CC core clusters. Each line represents a different cluster and
links together its values for each parameter (see text for details). Note
that the \citetalias{ascasibar08} model $\alpha$ parameter has been logged
for display purposes.}
\end{figure}

With a parametric analysis of a statistical sample we are able to address
the variation in cluster structure and, in particular, the comparison
between clusters with and without a cool core. In
Fig.~\ref{fig:jfpars_parallel} we explore the relationships between the
model parameters using a parallel coordinates plot, separating the CC and
non-CC clusters. This is a tool for hyperdimensional data visualization
\citep[e.g.][]{wegman90}, which is implemented in the \textsc{lattice} 
package in \Rproject\footnote{http://www.r-project.org} \citep{sarkar08}. A
series of variables (in this case the values of the best-fitting model
parameters) are plotted in 1 dimensional form along a common horizontal
axis spanning the range of the data in each case. These univariate plots
are arranged vertically and lines are drawn linking points for the same
cluster between all the variables.

It is clear from Fig.~\ref{fig:jfpars_parallel} that the CC cluster lines
are more tightly bunched compared to the non-CC clusters, demonstrating
that CC clusters have more uniform properties, possibly reflecting a more
diverse history for non-CC clusters. This could be because the formation of
a cool core only takes place in more relaxed clusters, which therefore
exhibit greater uniformity or could be the result of a dynamical
disturbance inhibiting or disrupting the formation of a cool core
\citep[cf.][]{san09b}. A much larger (but still representative) cluster
sample would be required to clarify these trends. Furthermore, this
systematic variation in structure between CC and non-CC clusters is
something that cosmological simulations should be able to reproduce.

%%%%%%%%%%%%%%%%%%%%%%%%%%%%%%%%%%%%%%%%%%%%%%%%%%%%%%%%%%%%%%%%%%%%
\section{Discussion \& Conclusions}
We have demonstrated that the \citetalias{ascasibar08} cluster model is
well suited to parametrizing the gas and mass distributions of a wide range
of galaxy clusters. The model is stable and easy to fit, and it performs
well with even sparse and noisy data (as few as $\sim$5 radial bins). It
also has the advantage of avoiding unphysical behaviour (i.e.  negative
values) and, being based on the \citet{hernquist90} profile, yields
convergent projected masses, facilitating easy comparison with
gravitational lensing measurements \citep[e.g.][]{richard09}. The use of
bootstrap resampling of the input gas temperature and density profile data
allows error estimates to be made directly on model parameters and any
derived quantity.

The fact that the model can be successfully applied to relatively poor
quality data greatly increases the number of clusters for which a full
X-ray mass analysis is possible, enabling better calibration of mass
proxies and scaling relations for cosmological studies. Furthermore, the
ability to easily determine important diagnostics of cluster physics
directly from the model is of great benefit to studies of galaxy feedback
and cluster evolution \citep[e.g.][]{san09b}. For example, the use of a
simple cluster modelling analysis like this is potentially a powerful tool
for exploring the effectiveness of baryon physics implementations in
cosmological simulations, by probing structural variations within synthetic
cluster populations and comparing them with real clusters.

%%%%%%%%%%%%%%%%%%%%%%%%%%%%%%%%%%%%%%%%%%%%%%%%%%%%%%%%%%%%%%%%%%%%
\section*{Acknowledgments}
AJRS acknowledges support from STFC. This work has made extensive use of
the \Rproject\ software package.

%%%%%%%%%%%%%%%%%%%%%%%%%%%%%%%%%%%%%%%%%%%%%%%%%%%%%%%%%%%%%%%%%%%%
\bibliography{/data/ajrs/stuff/latex/ajrs_bibtex}

%%%%%%%%%%%%%%%%%%%%%%%%%%%%%%%%%%%%%%%%%%%%%%%%%%%%%%%%%%%%%%%%%%%%
\appendix

\section{Estimating real variance}
\label{sec:appendix}
Consider some data $d_i$, with statistical errors $\sigma_i$ and a set of
estimates $\hat{d_i}$ for these values, derived from a model. The
deviations of $d_i$ from $\hat{d_i}$ consist of two parts, a statistical
scatter, $n_i$, which is Gaussian distributed with zero mean and standard
deviation $\sigma_i$, and a systematic offset $s_i$, which we are interested
in quantifying. Now
\begin{equation}
d_i = \hat{d_i} + n_i + s_i ,
\end{equation}
and we require an estimate of the fractional systematic deviation,
$s/\hat{d}$, allowing for the possibility that it may be zero, i.e. that
the fluctuations of $d_i$ about $\hat{d_i}$ are entirely due to the
measurement errors. It follows that
\begin{displaymath}
\left< (d_i - \hat{d_i})^2 \right> = \left< (n_i + s_i)^2 \right> = 
 \left< n_i^2 + s_i^2 +2 n_i s_i \right>
\end{displaymath}
\begin{displaymath}
= \left< n_i^2 \right> + \left< s_i^2 \right> = \sigma_i^2 + \left< s_i^2 \right> ,
\end{displaymath}
since $n_i s_i\rightarrow0$ if the measurement errors have zero mean and
are uncorrelated with $s_i$. So, our best estimate of $s_i^2$ is
\begin{equation}
\hat{s_i^2} = (d_i - \hat{d_i})^2 - \sigma_i^2 \;\; \mathrm{for} \; \hat{s_i^2} > 0 ,
\label{eqn:s}
\end{equation}
\begin{displaymath}
\hat{s_i^2}  = 0 \;\; \mathrm{for} \; \hat{s_i^2} < 0
\end{displaymath}

We can estimate $s/\hat{d}$ using
\begin{displaymath}
\left< \frac{s}{\hat{d}} \right> = \frac{1}{n} \sum_{i=1}^{n} \left< \frac{s_i}{\hat{d_i}} \right>
 = \frac{1}{n} \sum_{i=1}^{n} \frac{\left< s_i \right>}{\hat{d_i}}
\end{displaymath}
and insert the root mean square estimate for $\left< s_i \right>$. Therefore 
our estimate of $s/\hat{d}$, the fractional non-statistical deviation is 
given by
\begin{equation}
f = \frac{1}{n} \sum_{i=1}^{n} \frac{\sqrt{\hat{s_i^2}}}{\hat{d_i}}
\label{eqn:f.mean}
\end{equation}

\label{lastpage}

\end{document}